\documentclass{svjour2}                    
\smartqed  
\usepackage{graphicx}
\usepackage{amsfonts}
\usepackage{amssymb}
\begin{document}

\title{Oscillatory behaviour of closed isotropic models in second order gravity theory}

\author{J. Miritzis}

\institute{J. Miritzis \at Department of Marine Sciences,
University of the Aegean, University Hill, Mytilene 81100,
Greece\\ \email{imyr@aegean.gr}}

\date{Received: date / Accepted: date}
\maketitle

\begin{abstract}
Homogeneous and isotropic models are studied in the Jordan frame
of the second order gravity theory. The late time evolution of the
models is analysed with the methods of the dynamical systems. The
normal form of the dynamical system has periodic solutions for a
large set of initial conditions. This implies that an initially
expanding closed isotropic universe may exhibit oscillatory
behaviour.


\end{abstract}

\section{Introduction}

Quadratic gravitational Lagrangians were proposed shortly after
the formulation of general relativity (GR) as alternatives to
Einstein's theory. Gravity modifications in the form of
higher-order curvature invariants in the Lagrangian are generally
known as higher-order gravity (HOG) theories. They arise in
string-theoretic considerations, e.g., brane models with
Gauss-Bonett terms \cite{lno} or models with a scalar field
coupled to the Gauss-Bonett invariant \cite{nos} (see \cite{gave}
for a review) and generally involve linear combinations of all
possible second order invariants that can be formed from the
Riemann, Ricci and scalar curvatures. A quarter of a century ago
there was a resurgence of interest in such theories in an effort
to explain inflation. The reasons for considering HOG theories
were multiple. Firstly, it was hoped that higher order Lagrangians
would create a first approximation to quantum gravity, due to
their better renormalisation properties than GR \cite{stel}.
Secondly, it was reasonable to expect that on approach to a
spacetime singularity, curvature invariants of all orders ought to
play an important dynamical role. Far from the singularity, when
higher order corrections become negligible, one should recover GR.
Furthermore, it was hoped that these generalized theories of
gravity might exhibit better behavior near singularities. Thirdly,
inflation emerges in these theories in a most direct way. In one
of the first inflationary models, proposed in 1980 by Starobinsky
\cite{star}, inflation is due to the $R^{2}$ correction term in a
gravitational Lagrangian $L=R+\beta R^{2}$ where $\beta$ is a
constant.

Recently there is a revival of interest in HOG theories in an effort to
explain the accelerating expansion of the Universe \cite{cdtt,cct2}. The
general idea is to add an $1/R$ term to the Einstein-Hilbert Lagrangian or
more generally to consider $R+\alpha R^{-n}$ Lagrangians \cite{caretal,nood}.
As the Universe expands, one expects that the inverse curvature terms will
dominate and produce the late time accelerating expansion. Most studies are
restricted to simple Friedmann-Robertson-Walker (FRW) models because of the
complexity of the field equations. At present, the observational viability of
these models is a subject of active research (see \cite{cno,amen1,amen2,cafr}
and references therein). However, it seems that a large class of Lagrangians
may fit the observational data, but simple models based on the FRW metric are
insufficient to pick the correct Lagrangian (see for example \cite{cct1} for
the reconstruction of the $f(R)$ theory which best reproduces the observed
cosmological data). For more general spacetimes it may even be meaningless to
say that $R$ is small in some epoch of cosmic evolution and large in some
other one (for a thorough critic see \cite{soko}).

In this paper we investigate the late time evolution of flat and positively
curved FRW models with a perfect fluid in the $R+\beta R^{2}$ theory. This is
the simplest generalization of the Einstein-Hilbert Lagrangian and the
addition of the quadratic term represents a correction to general relativity.
The simple vacuum case was studied in \cite{miri1}, where oscillatory
behaviour of the solutions of closed models was found. Since HOG theories in
vacuum are conformally equivalent to GR with a scalar field, it is tempting to
say that the $R^{2}$ contribution has predictable cosmological consequences
\cite{bahe}. However, this\ is an oversimplification of the picture (see
\cite{cno,nood1} for specific examples). The two frames are mathematically
equivalent, but physically they provide different theories. In the Jordan
frame, gravity is described entirely by the metric $g_{\mu\nu}.$ In the
Einstein frame, the scalar field exhibits a non-metric\ aspect of the
gravitational interaction, reflecting the additional degree of freedom due to
the higher order of the field equations in the Jordan frame. Inclusion of
additional matter fields, further complicates the situation and while the
field equations in the Einstein frame are formally the Einstein equations,
nevertheless this theory is not physically equivalent to GR. There is no
universally acceptable answer to the issue \textquotedblleft which conformal
frame is physical\textquotedblright\ \cite{bran} (see \cite{maso} for a
thorough analysis of different views).

The plan of the paper is as follows. Next Section contains a short comment on
the stability of well-known power-law solutions. The field equations are
written as a constrained four-dimensional polynomial dynamical system. Section
3 contains the analysis of the flat case. The so-called normal form of the
dynamical system greatly simplifies the problem, since two of the equations
decouple. In Section 4 we study the qualitative behaviour of the solutions
near the equilibrium points of positively curved models and analyse their late
time evolution. It is shown that an initially expanding closed FRW universe
may exhibit oscillatory behaviour.

\section{Field equations}

The general gravitational Lagrangian in four-dimensional
spacetimes contains curvature invariants of all orders,
$R,R^{2},R_{\mu\nu}R^{\mu\nu},...$. The term
\linebreak$R_{\mu\nu\rho\sigma}R^{\mu\nu\rho\sigma}$ is omitted
because of the Gauss-Bonnet theorem. A further simplification can
be done in homogeneous and isotropic spacetimes where the
variation of $R_{\mu\nu}R^{\mu\nu}$ with respect to the metric is
proportional to the variation of $R^{2}$ \cite{baot}. We conclude
that for isotropic cosmologies the gravitational Lagrangian
contains only powers of the scalar curvature and we may consider
HOG theories derived from Lagrangians of the form
\[
L=f\left(  R\right)  \sqrt{-g}+L_{\mathrm{matter}},
\]
where $f$ is an arbitrary smooth function. It is well-known that the
corresponding field equations are fourth-order and take the form
\begin{equation}
f^{\prime}\left(  R\right)  R_{\mu\nu}-\frac{1}{2}f\left(  R\right)  g_{\mu
\nu}-\nabla_{\mu}\nabla_{\nu}f^{\prime}\left(  R\right)  +g_{\mu\nu}\Box
f^{\prime}\left(  R\right)  =T_{\mu\nu},\label{hoge}%
\end{equation}
where $\Box=g^{\alpha\beta}\nabla_{\alpha}\nabla_{\beta}$ and a
prime ($^{\prime}$) denotes differentiation with respect to $R.$
The generalised Bianchi identities imply that
$\nabla^{\mu}T_{\mu\nu}=0.$ Contraction of (\ref{hoge}) yields the
trace equation
\begin{equation}
3\Box f^{\prime}\left(  R\right)  +f^{\prime}\left(  R\right)  R-2f\left(
R\right)  =T.\label{trace}%
\end{equation}
In contrast to GR where the relation of $R$ to $T$ is algebraic, in HOG
theories the trace equation (\ref{trace}) is a differential equation for $R,$
with $T$ as source term \cite{olmo}. This suggests that in HOG theories both
the metric and the scalar curvature are dynamical fields.

In the following, we consider a quadratic Lagrangian without cosmological
constant, i.e. $f\left(  R\right)  =R+\beta R^{2},\ \beta>0,$ and confine our
attention to cosmologies with a perfect fluid with energy density $\rho$ and
pressure $p,$ of the form
\[
p=(\gamma-1)\rho,\ \ \ \ 0\leq\gamma\leq2.
\]
For homogeneous and isotropic spacetimes\footnote{We adopt the metric and
curvature conventions of \cite{wael}. Here, $a\left(  t\right)  $ is the scale
factor, an overdot denotes differentiation with respect to time $t,$ and units
have been chosen so that $c=1=8\pi G.$} described by the standard FRW metric
we need the following useful relations ($i,j=1,2,3$ and $k=0,\pm1$)
\begin{equation}
R_{00}=-3\frac{\ddot{a}}{a},\ R_{ij}=\left(  \frac{\ddot{a}}{a}+2\frac{\dot
{a}^{2}}{a^{2}}+2\frac{k}{a^{2}}\right)  g_{ij},\ \ R=6\left(  \frac{\ddot{a}%
}{a}+\frac{\dot{a}^{2}}{a^{2}}+\frac{k}{a^{2}}\right)  . \label{isot}%
\end{equation}
The $00$ component of (\ref{hoge}) is
\begin{equation}
H^{2}+\frac{k}{a^{2}}+2\beta\left[  R\left(  H^{2}+\frac{k}{a^{2}}\right)
+H\dot{R}-\frac{R^{2}}{12}\right]  =\frac{1}{3}\rho, \label{0-0}%
\end{equation}
where $H=\dot{a}/a,$ is the Hubble function.

At this point we make a digression. For flat, $k=0,$ models, differentiating
the relation $R=6\dot{H}+12H^{2}$ (which comes from the third of (\ref{isot}))
with respect to $t$, equation (\ref{0-0}) takes the form%
\begin{equation}
H^{2}+6\beta\left(  2H\ddot{H}+6H^{2}\dot{H}-\dot{H}^{2}\right)  =\frac{1}%
{3}\rho. \label{00flat}%
\end{equation}
For radiation, $\gamma=4/3$, there exists a special solution $a\left(
t\right)  =t^{1/2}$ as $t\rightarrow0.$ As Barrow and Middleton point out
\cite{bami}, this is also an exact vacuum solution of the purely quadratic
theory, in the sense that it solves $2H\ddot{H}+6H^{2}\dot{H}-\dot{H}^{2}=0.$
However, this solution is unstable as we shall see in a moment (see
\cite{cofl} for a detailed stability analysis of isotropic models in general
$f\left(  R\right)  $ theories). Following Carroll et al \cite{caretal}, we
reduce the order of (\ref{00flat}) in vacuum by defining%
\[
X=-H,\ Y=\dot{H}\ \Rightarrow\ \ddot{H}=-Y\frac{dY}{dX}.
\]
Then, the asymptotic values of the function $U\left(  X\right)  =-X^{2}/Y$ as
$X\rightarrow0,$ correspond to the exponents $p$ for power-law solutions
$a\left(  t\right)  =t^{p}$. We apply this technique to (\ref{00flat}) and we
find the first-order equation%
\begin{equation}
\frac{dU}{dX}=\frac{1}{12\beta}\frac{U^{3}}{X^{3}}-3\frac{U}{X}\left(
U-\frac{1}{2}\right)  , \label{redu}%
\end{equation}
with the corresponding direction field shown in Fig. 1. We see that the
solution $U\left(  X\right)  =1/2$ is a past attractor ($X\rightarrow-\infty
$), but becomes unstable as $X\rightarrow0,$ in agreement with \cite{cofl},
and that $\left\vert U\left(  X\right)  \right\vert \rightarrow0,$
corresponding to the singularity $\left\vert \dot{H}\right\vert \rightarrow
\infty$. This is also evident by studying the asymptotic behaviour of
solutions of the linearised equation near the constant solutions $1/2$ and
$0.$ \begin{figure}[h]
\begin{center}
\includegraphics[scale=0.7]{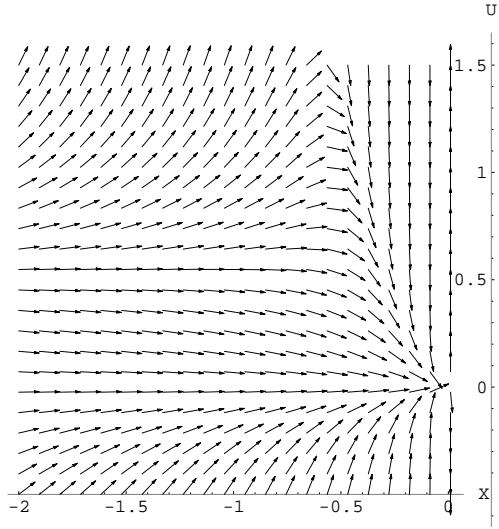}
\end{center}
\caption{Direction field of the reduced equation (\ref{redu})}%
\label{fig1}%
\end{figure}

We now continue our discussion about the choice of variables. Setting $x=1/a,$
the $00$ equation becomes
\begin{equation}
H^{2}+kx^{2}+2\beta\left[  R\left(  H^{2}+kx^{2}\right)  +H\dot{R}-\frac
{R^{2}}{12}\right]  =\frac{1}{3}\rho, \label{00}%
\end{equation}
and the evolution equation for $x$ is
\begin{equation}
\dot{x}=-xH. \label{chi}%
\end{equation}
The evolution equation for $H$ comes from the third of (\ref{isot}) and takes
the form%
\begin{equation}
\dot{H}=\frac{1}{6}R-2H^{2}-kx^{2}. \label{hubb}%
\end{equation}
The conservation equation%
\begin{equation}
\dot{\rho}=-3\gamma\rho H, \label{cons}%
\end{equation}
is a consequence of the Bianchi identities. With the relation $\Box
R=-\ddot{R}-3H\dot{R},$ equation (\ref{trace}) becomes
\begin{equation}
\ddot{R}+3H\dot{R}+\frac{1}{6\beta}R=\frac{1}{6\beta}\left(  4-3\gamma\right)
\rho. \label{trac}%
\end{equation}
This equation is usually considered as superfluous, since it follows from the
differentiation of the equation (\ref{00}) with respect to $t$; for example
equation (\ref{trac}) was used in \cite{berk} as a control of the accuracy of
numerical investigations of Bianchi type I and IX models. Thus, one can chose
$R,x,H~$and $\rho$ as dynamical variables which obey the evolution equations
(\ref{00}), (\ref{chi}), (\ref{hubb}) and (\ref{cons}) and constitute a
four-dimensional dynamical system.

The form of equation (\ref{00}) suggest the choice of expansion normalized
variables of the type
\begin{equation}
u\sim R/H,~\Omega\sim\rho/H^{2},... \label{tran}%
\end{equation}
Detailed studies using this approach for general $f\left(  R\right)  $
theories can be found in \cite{amen2,ctd}. However, even in the simplest case
of flat models in vacuum, numerical investigation of the system with initial
values $H\left(  0\right)  >0$ shows that $H\left(  t\right)  $ exhibit damped
oscillations, with almost zero minima. Therefore although permissible, the
transformation (\ref{tran}) may induce fake singularities to the solutions.
Moreover, as mentioned in \cite{wael}, a drawback of this choice of variables
is that it does not give a complete description of the evolution for bouncing
or recollapsing models. If the Hubble parameter passes through zero the
logarithmic time coordinate is ill-defined and the transformation (\ref{tran})
is singular (see \cite{gld} for a comparison of compact and non-compact variables).

In order to circumvent these difficulties, we introduce one more degree of
freedom by using (\ref{trac}), thus augmenting the dimension of the dynamical
system. The state $(R,\dot{R},x,\rho,H)$ of the system lies on the
hypersurface of $\mathbb{R}^{5}$ defined by the constraint (\ref{00}). The
presence of $(R,\dot{R})$ in the state vector reflects the fact that there are
additional degrees of freedom in HOG theories than in GR (cf. the remark after
equation (\ref{trace})).

We define $x_{1}=R,$ $x_{2}=\dot{R}$ and use (\ref{00}) to eliminate $\rho,$
so that equations (\ref{trac}), (\ref{chi}) and (\ref{hubb}) constitute a
four-dimensional system. The parameter $\beta$ may be used to define
dimensionless variables by rescaling%
\[
x_{1}\rightarrow\frac{2}{\beta}x_{1},\ x_{2}\rightarrow\sqrt{\frac{2}%
{3\beta^{3}}}~x_{2},\ H\rightarrow\frac{H}{\sqrt{6\beta}},\ x\rightarrow
\frac{x}{\sqrt{6\beta}},\ t\rightarrow\sqrt{6\beta}~t,
\]
and our system becomes%
\begin{eqnarray}
\dot{x}_{1}&=&x_{2},\nonumber\\ \dot{x}_{2}&=&-x_{1}+\left(
1-3\gamma\right)  x_{2}H+\frac{\left( 4-3\gamma\right)
}{4}Z,\label{full}\\ \dot{x}&=&-xH,\nonumber\\
\dot{H}&=&2x_{1}-2H^{2}-kx^{2},\nonumber
\end{eqnarray}
with $Z=\left[  H^{2}+kx^{2}-4x_{1}^{2}+4x_{1}\left(  H^{2}+kx^{2}\right)
\right]  $.

\textbf{Remark.} The system (\ref{full}) is not an arbitrary \textquotedblleft
free\textquotedblright\ four-dimensional system. In view of (\ref{00}) the
initial conditions have to satisfy the condition
\begin{equation}
H^{2}+kx^{2}+4x_{1}\left(  H^{2}+kx^{2}\right)  +4Hx_{2}-4x_{1}^{2}%
\geq0.\label{ineq}%
\end{equation}
With a little manipulation of the equations (\ref{full}) it can be shown that,
once we start with initial conditions satisfying (\ref{ineq}) at time $t_{0}$,
the solutions of the system satisfy this inequality for all $t>t_{0}.$ Thus
the field equations share the general property of the Einstein equations,
namely that the subsequent evolution of the system is such that the solutions
respect the constraint.

\section{Flat models}

In the flat, ($k=0$), case the dimension of the dynamical system
is reduced by one, since the evolution equation for $x$ decouples
from the remaining
equations. The corresponding system is%
\begin{eqnarray}
\dot{x}_{1}&=&x_{2},\nonumber\\ \dot{x}_{2}&=&-x_{1}+\left(
1-3\gamma\right)  x_{2}H-\left( 4-3\gamma\right)
x_{1}^{2}+\frac{\left(  4-3\gamma\right)  }{4}H^{2}+\left(
4-3\gamma\right)  x_{1}H^{2},\nonumber\\
\dot{H}&=&2x_{1}-2H^{2},\label{flat}
\end{eqnarray}
i.e., the vector field does not depend on the dynamical variable
$x.$ Vacuum models are significantly simpler to analyse, but we do
not study them separately as they arise formally by setting
$\gamma=4/3$ in all the equations, while (\ref{ineq}) describing
the phase space becomes equality. The only equilibrium point of
(\ref{flat}) is the origin and corresponds to flat empty
spacetime. The eigenvalues of the Jacobian matrix at the origin
are $\pm i,0$ and therefore we cannot infer about stability using
the linearisation theorem. For nonhyperbolic equilibrium points
there exist no general methods for studying their stability. The
normal form of the system may provide some information about the
behaviour of the solutions near the equilibrium. The normal form
theory consists in a nonlinear coordinate transformation that
allows to simplify the nonlinear part of the system (cf.
\cite{perko} for a brief introduction). This task will be
accomplished in three steps in some detail for the convenience of
readers with no previous knowledge of the method.

1. Let $P$ be the matrix formed from the eigenvectors which
transforms the linear part of the vector field into Jordan
canonical form. We write
(\ref{flat}) in vector notation (with $\mathbf{x}=\left(  x_{1},x_{2}%
,H\right)  $) as
\begin{equation}
\mathbf{\dot{x}}=A\mathbf{x}+\mathbf{F}\left(  \mathbf{x}\right)
,
\label{sys3a}%
\end{equation}
where $A$ is the linear part of the vector field and
$\mathbf{F}\left( \mathbf{0}\right)  =\mathbf{0}$.

2. Using the matrix $P$, we define new variables, $\left(  y_{1}%
,y_{2},y\right)  \equiv\mathbf{y}$, by the equations%
\[
y_{1}=x_{1},\ \ y_{2}=-x_{2},\ \ \ \ y=H+2x_{2},
\]
or in vector notation $\mathbf{y}=P\mathbf{x},$ so that
(\ref{sys3a}) becomes
\[
\mathbf{\dot{y}}=P^{-1}AP\mathbf{y}+P^{-1}\mathbf{F}\left(  P\mathbf{y}%
\right)  .
\]
Denoting the canonical form of $A$ by $B$ we finally obtain the
system
\begin{equation}
\mathbf{\dot{y}}=B\mathbf{y}+\mathbf{f}\left(  \mathbf{y}\right)
,
\label{sys3b}%
\end{equation}
where $\mathbf{f}\left(  \mathbf{y}\right)
:=P^{-1}\mathbf{F}\left(
P\mathbf{y}\right)  .$ In components system (\ref{sys3b}) is%
\begin{eqnarray*}
&&\left[
\begin{array}
[c]{c}%
\dot{y}_{1}\\ \dot{y}_{2}\\
\dot{y}%
\end{array}
\right]=\left[
\begin{array}
[c]{ccc}%
0 & -1 & 0\\ 1 & 0 & 0\\ 0 & 0 & 0
\end{array}
\right]  \left[
\begin{array}
[c]{c}%
y_{1}\\ y_{2}\\ y
\end{array}
\right]+ \\ &&\left[
\begin{array}
[c]{c}%
0\\
\left(  4-3\gamma\right)  y_{1}^{2}-\left(  3\gamma+2\right)  y_{2}^{2}%
-\frac{4-3\gamma}{4}y^{2}-3yy_{2}-\left(  4-3\gamma\right)
y_{1}\left( 2y_{2}+y\right)  ^{2}\\
-2\left(  4-3\gamma\right)  y_{1}^{2}+2\left(  3\gamma-2\right)  y_{2}%
^{2}-\frac{3\gamma}{2}y^{2}-2yy_{2}+2\left(  4-3\gamma\right)
y_{1}\left(
2y_{2}+y\right)  ^{2}%
\end{array}
\right]  .
\end{eqnarray*}
Inequality (\ref{ineq}) imposes the constraint
\begin{equation}
y^{2}-4\left(  y_{1}^{2}+y_{2}^{2}\right)  +4y_{1}\left(
y+2y_{2}\right)
^{2}\geq0. \label{ineq1}%
\end{equation}
3. Under the non-linear change of variables%
\begin{eqnarray}
y_{1}&\rightarrow&y_{1}+3\gamma y_{1}^{2}+\left(  3\gamma-2\right)
y_{2}^{2}+\frac{4-3\gamma}{4}y^{2}+\frac{3}{4}y_{2}y,\nonumber\\
y_{2}&\rightarrow&
y_{2}+4y_{1}y_{2}+\frac{3}{4}y_{1}y,\label{nonl}\\ y&\rightarrow&
y-2y_{1}y_{2}+2y_{1}y,\nonumber
\end{eqnarray}
and keeping only terms up to second order, the system transforms to%

\begin{eqnarray}
\dot{y}_{1}&=&-y_{2}-\frac{3}{2}y_{1}y+\mathcal{O}\left( 3\right)
,\nonumber\\
\dot{y}_{2}&=&y_{1}-\frac{3}{2}y_{2}y+\mathcal{O}\left(  3\right)
,\nonumber\\ \dot{y}&=&6\left(  \gamma-1\right)  \left(
y_{1}^{2}+y_{2}^{2}\right)
-\frac{3\gamma}{2}y^{2}+\mathcal{O}\left(  3\right)  .\nonumber
\end{eqnarray}
Finally, defining cylindrical coordinates $\left(
y_{1}=r\cos\theta
,y_{2}=r\sin\theta,y=y\right)  ,$ we obtain%
\begin{eqnarray}
\dot{r}&=&-\frac{3}{2}ry+\mathcal{O}\left(  3\right) ,\nonumber\\
\dot{\theta}&=&1+\mathcal{O}\left(  2\right) ,\label{cylifl}\\
\dot{y}&=&6\left(  \gamma-1\right)  r^{2}-\frac{3\gamma}{2}y^{2}%
+\mathcal{O}\left(  3\right)  .\nonumber
\end{eqnarray}
We may continue to simplify the third order terms and the result
should be
\begin{eqnarray*}
\dot{r}&=&a_{1}ry+a_{2}r^{3}+a_{3}ry^{2}+\mathcal{O}\left(
4\right)  ,\\ \dot{\theta}&=&1+\mathcal{O}\left(  2\right) ,\\
\dot{y}&=&b_{1}r^{2}+b_{2}y^{2}+b_{3}r^{2}y+b_{4}y^{3}+\mathcal{O}\left(
4\right)  .
\end{eqnarray*}
This is the normal form in cylindrical coordinates of every
three-dimensional vector field with linear part
\[
\left[
\begin{array}
[c]{ccc}%
0 & -1 & 0\\ 1 & 0 & 0\\ 0 & 0 & 0
\end{array}
\right]  \left[
\begin{array}
[c]{c}%
y_{1}\\ y_{2}\\ y
\end{array}
\right]  ,
\]
(see \cite{guho} p. 377). However, since we are interested on the
behaviour of the solutions only near the origin, we truncate the
vector field at $\mathcal{O}\left(  2\right)  .$ We note that the
$\theta$ dependence of the vector field has been eliminated, so
that we can study the system in the $(r,y)$ space. The second
equation of (\ref{cylifl}) implies that the trajectory in the
$y_{1}-y_{2}$ plane spirals with angular velocity $1$. The
projection of (\ref{cylifl}) on the $r-y$ plane is
\begin{eqnarray}
\dot{r}&=&-\frac{3}{2}ry,\;\nonumber\\ \dot{y}&=&6\left(
\gamma-1\right)  r^{2}-\frac{3\gamma}{2}y^{2}.
\label{2dfl}%
\end{eqnarray}
This system belongs to a family of systems studied in 1974 by
Takens \cite{takens} (see also \cite{guho} for a description of
all phase portraits for different values of the parameters).

System (\ref{2dfl}) is invariant under the transformation
$t\rightarrow-t,$ $y\rightarrow-y$ (which implies that all
trajectories are symmetric with respect to the $r$ axis) and the
line $r=0$ is invariant. Note also that the system (\ref{2dfl})
has invariant lines $y=\pm2r$ (compare with the similar system in
\cite{miri2}). The behaviour of the solutions depends on the
parameter $\gamma$ and as we shall see, $\gamma=1$ is a
bifurcation value.

\begin{figure}[h]
\begin{center}
\includegraphics[scale=0.7]{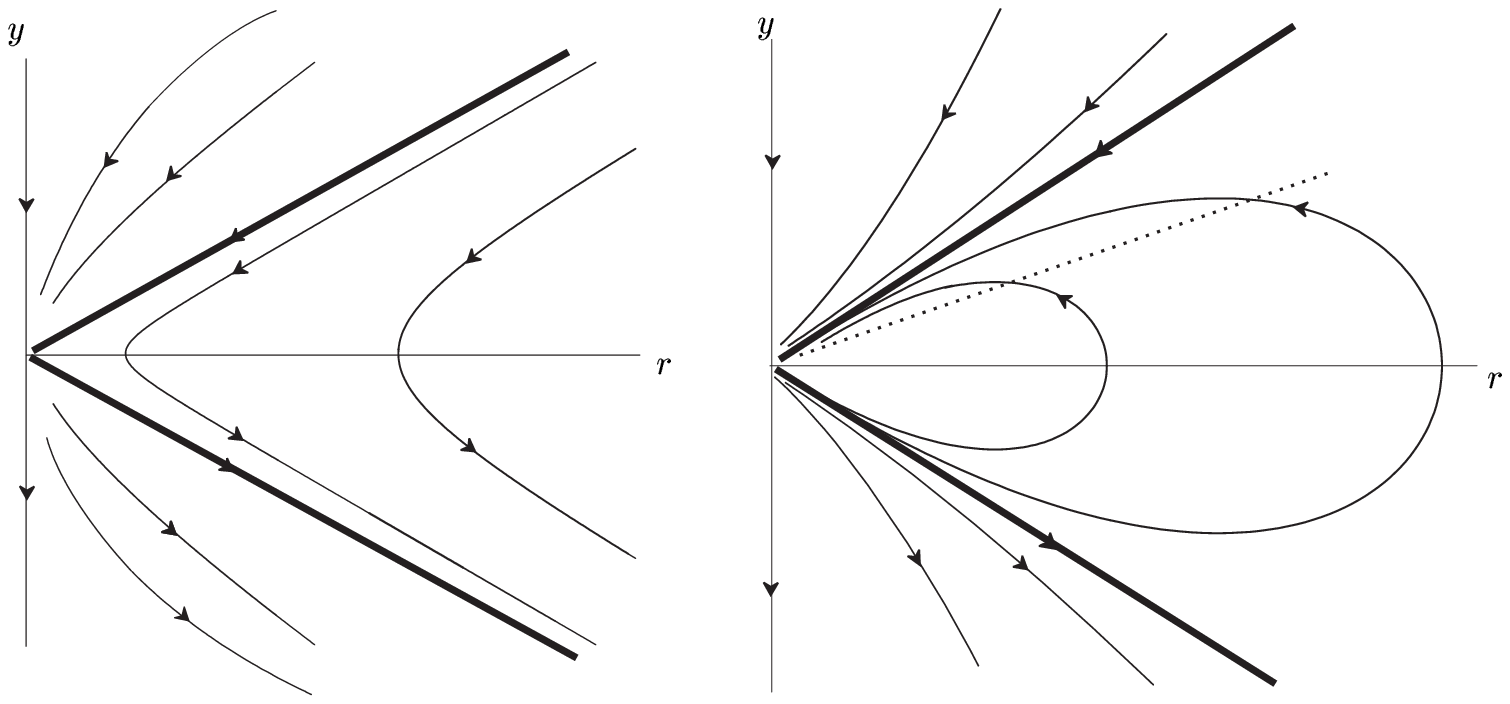}
\end{center}
\caption{Phase portrait of (\ref{2dfl}) for $\gamma<1$ and
$\gamma>1$. The invariant lines $y=\pm2r$ are shown with heavy
lines and $y=2\sqrt{\gamma
-1}~r$ by the dotted line.}%
\label{fig2}%
\end{figure}

Case I, $\gamma<1.$ We observe that $y$ is always decreasing along
the orbits while $r$ is decreasing in the first quadrant. Since no
trajectory can cross the line $y=2r,$ all trajectories starting
above this line, approach the origin asymptotically. The phase
space of the dynamical system (\ref{2dfl}) is not the whole $r-y$
plane, because of the constraint (\ref{ineq1}). In terms of the
variables (\ref{nonl}) and neglecting fourth-order terms the
constraint
becomes%
\begin{equation}
\left(  y^{2}-4r^{2}\right)  (1+6\gamma y_{1})\geq0. \label{ineq2}%
\end{equation}
Since we are interested on trajectories starting close to the
origin, for initial values of $y_{1}$ satisfying $\left\vert
y_{1}\left(  0\right) \right\vert \leq1/6\gamma$ the first of
(\ref{2dfl}) guarantees that $\left\vert y_{1}\left(  t\right)
\right\vert $ remains less than $1/6\gamma $, so that
(\ref{ineq2}) implies
\[
y^{2}-4r^{2}\geq0.
\]
Therefore, we should consider only trajectories starting above the
line $y=2r$ and according to the previous discussion all these
trajectories asymptotically approach the origin.

Case II, $\gamma>1.$ In the first quadrant $r$ is decreasing along
the orbits and, that $\dot{y}$ vanishes along the line
$y=2\sqrt{\gamma-1}~r.$ Once a trajectory crosses the line
$y=2\sqrt{\gamma-1}~r,$ it is trapped between the lines
$y=2\sqrt{\gamma-1}~r$ and $y=2r,$ and since $\dot{r}<0,$ it
approaches the origin asymptotically.

Case III, $\gamma=1.$ It is evident that all trajectories are
straight lines approaching asymptotically the origin.

We conclude that the late time behaviour of flat models is similar
to the future predicted by GR. More precisely, all initially
expanding flat models close to the state $R=\dot{R}=H=\rho=0,$
asymptotically approach the flat empty spacetime.

\section{Positively curved models}

In this section we consider an initially expanding closed universe
described by the full four-dimensional system (\ref{full}) with
$k=1$. There are two equilibria:

$\mathcal{M}:x_{1}=0,x_{2}=0,x=0,H=0.$ This corresponds to the
limiting state of an almost empty, slowly varying universe with
$H\rightarrow0,$ while the scale factor goes to infinity. The
point $\mathcal{M}$ which resembles to the Minkowski solution, is
located at the boundary of the phase space.

$\mathcal{S}:H=0,x_{2}=0,x=\bar{x}\equiv\sqrt{\frac{2-3\gamma}{3\gamma-4}%
},x_{1}=\bar{x}_{1}\equiv\bar{x}^{2}/2,$ with $2/3<\gamma<4/3.$ It
is a static
solution and the eigenvalues of the Jacobian matrix at $\mathcal{S}$ are%
\[
\pm\frac{8-9\gamma\pm\sqrt{3\gamma\left(  36\gamma^{2}-69\gamma+32\right)  }%
}{2\left(  3\gamma-4\right)  }.
\]
The real parts of the eigenvalues are nonzero for almost
all\footnote{More precisely, the real parts of the eigenvalues are
zero for
\[
\frac{2}{3}<\gamma\leq\frac{23-\sqrt{17}}{24}\approx0.79,
\]
and nonzero in the rest of the interval
$(\frac{2}{3},\frac{4}{3}).$} permissible values of $\gamma$, and
we conclude that the local stable and unstable manifolds through
$\mathcal{S}$ are both two-dimensional. The point $\mathcal{S}$
corresponds to the Einstein static universe, where the effective
cosmological constant is provided by the curvature equilibrium
$\bar{x}_{1}.$ To see this, it is sufficient to write equation
(\ref{00}) in the original variables at the equilibrium point as
\[
k\bar{x}^{2}+\frac{\Lambda}{3}=\frac{1}{3}\bar{\rho},
\]
with $\Lambda>0.$ The cosmological constant depends on both the
parameters $\beta$ and $\gamma,$ for example, for $\gamma=1,$
$\Lambda=\left( 12\beta-1\right)  ^{2}/2\beta.$ Static solutions
have little interest as cosmological models and we turn our
attention to the other equilibrium, $\mathcal{M}$.

The point $\mathcal{M}$ is a nonhyperbolic equilibrium and we find
again the normal form of the system, which is given by
(\ref{normalf}) in the Appendix. Defining cylindrical coordinates
$\left(  y_{1}=r\cos\theta,y_{2}=r\sin
\theta,x=x,y=y\right)  ,$ we obtain%
\begin{eqnarray}
\dot{r}&=&-\frac{3}{2}ry+\mathcal{O}\left(  3\right) ,\nonumber\\
\dot{\theta}&=&1+\mathcal{O}\left(  2\right) ,\nonumber\\ \dot{x}
&=&-xy+\mathcal{O}\left(  3\right) ,\label{cyli}\\
\dot{y}&=&6\left(  \gamma-1\right)  r^{2}-\frac{3\gamma-2}{2}x^{2}%
-\frac{3\gamma}{2}y^{2}+\mathcal{O}\left(  3\right)  .\nonumber
\end{eqnarray}
We truncate the vector field at $\mathcal{O}\left(  2\right)  $
and we note again that the $\theta$ dependence of the vector field
has been eliminated, so that we can study the system in the
$(r,x,y)$ space. We write the first and
third of (\ref{cyli}) as a differential equation%
\[
\frac{dr}{dx}=\frac{3}{2}\frac{r}{x},
\]
which has the general solution%
\begin{equation}
r=Ax^{3/2},\ \ A>0. \label{r-equ}%
\end{equation}
Substitution of (\ref{r-equ}) into the fourth equation of
(\ref{cyli}) yields
the projection of the fourth-dimensional system on the $x-y$ plane, namely%
\begin{eqnarray}
\dot{x}&=&-xy,\nonumber\\
\dot{y}&=&b\left(  \gamma-1\right)  x^{3}-\frac{3\gamma-2}{2}x^{2}%
-\frac{3\gamma}{2}y^{2},\ \ \ b>0. \label{2dim}%
\end{eqnarray}

Some general properties of the solutions of (\ref{2dim}) follow by
inspection. By standard arguments all trajectories are symmetric
with respect to the $x$ axis. Note that the new $x$ defined by
(\ref{nonl1}) remains non-negative for initial values $y_{1}\left(
0\right)  $ sufficiently small and, since the line $x=0$ is
invariant, any trajectory starting at the half plane $x>0$ remains
there for all $t>0.$ System (\ref{2dim}) has two equilibrium
points, the origin $\left(  0,0\right)  $ and $\left(
x_{\ast},0\right)  $, where
\[
x_{\ast}=\frac{3\gamma-2}{2b\left(  \gamma-1\right)  },
\]
and therefore $x_{\ast}>0$ for $\gamma<2/3$ or $\gamma>1.$ The
origin is again a nonhyperbolic equilibrium point. Computation of
the Jacobian matrix at the equilibrium $\left(  x_{\ast},0\right)
$ shows that for the linearised system, this point is a saddle for
$0\leq\gamma<2/3,$ and a center for $1<\gamma\leq2$. For
$\gamma>1,$ it is easy to see that $x$ is decreasing in the first
quadrant and $y$ is decreasing along the orbits in the strip
$0<x<x_{\ast}$. On any orbit starting in the first quadrant with
$x<x_{\ast},$ $y$ becomes zero at some time and the trajectory
crosses vertically the $x-$axis. Once the trajectory enters the
second quadrant, $x$ increases and $y$ decreases. For
$2/3<\gamma\leq1$, all trajectories starting in the first quadrant
follow the same pattern.

System (\ref{2dim}) has a first integral, \textit{viz}.%
\begin{eqnarray*}
\phi\left(  x,y\right)&=&-\frac{2b}{3}x^{3\left( 1-\gamma\right)
}+x^{2-3\gamma}+\frac{y^{2}}{x^{3\gamma}},\ \ \gamma\neq1,\\
\phi\left(  x,y\right)&=&\frac{1}{x}+\frac{y^{2}}{x^{3}},\ \
\gamma=1.
\end{eqnarray*}
This can be seen by writing (\ref{2dim}) as
\[
\frac{dy}{dx}=\frac{bx^{3}-\frac{3\gamma-2}{2}x^{2}-\frac{3\gamma}{2}y^{2}%
}{-yx}.
\]
Setting $y^{2}=z,$ we obtain a linear differential equation for
$z$ which is easily integrable. The level curves of $\phi$ are the
trajectories of the system.

\medskip

We shall show for the system (\ref{2dim}) that: \emph{(i) for every }%
$\gamma\in(\frac{2}{3},2]$ \emph{there are no solutions
asymptotically approaching the origin (ii) for
}$\gamma\in(\frac{2}{3},1]$\emph{ there are no periodic solutions
and (iii) for }$\gamma\in(1,2]$ \emph{there exist periodic
solutions and the basin of every periodic trajectory is the set}
\[
\left\{  \left(  x,y\right)  \in\mathbb{R}^{2}:y^{2}+x^{2}-\frac{2b}{3}%
x^{3}<0,x>0\right\}  .
\]

\begin{proof}
The proof mimics that found in \cite{miri1} for the simple case
$\gamma=4/3.$

Let $\phi\left(  x,y\right)  =C$. We have%
\begin{equation}
y^{2}=x^{2}\left(  -1+\frac{2b}{3}x+Cx^{3\gamma-2}\right)  , \label{xy}%
\end{equation}
which implies that the function
\[
f\left(  x\right)  =-1+\frac{2b}{3}x+Cx^{3\gamma-2}%
\]
must be non-negative. We consider two cases.

1. $C\geq0$. Then $f$ is strictly increasing for $x\geq0$ and
$f\left( 0\right)  =-1$, thus $f$ has a unique root $x_{1}>0$
depending on $C$. It
follows that for $C\geq0$ any orbit starting in the first quadrant satisfies%
\[
x\geq x_{1}\left(  C\right)  >0,
\]
i.e., there are no solutions approaching the axis $x=0$. These
solutions are not closed since they intersect the $x-$axis only
once at $x_{1}(C)$.

2. $C<0$. If $2/3<\gamma\leq1$, then $f$ has again a unique root
hence the trajectories follow the same pattern as in case 1. If
$\gamma>1$, then $f$ has two zeros, say $x_{1}(C)<x_{2}(C)$. This
means that $0<x_{1}(C)<x<x_{2}(C)$, i.e., $x$ is bounded, and by
(\ref{xy}), so is $y$. Thus, an orbit of (\ref{2dim}) starting in
the first quadrant crosses the $x-$axis at $x_{1}\left(  C\right)
$ and re-enters in the first quadrant crossing the $x-$axis at
$x_{2}\left(  C\right)  $ i.e. it is a closed curve and represents
a periodic solution. The curve corresponding to $C=0$ separates
the phase space into two disjoint regions I and II. In region II,
($C<0$), every trajectory corresponds to a periodic solution and
we conclude that the basin of every periodic trajectory is the set
$y^{2}+x^{2}-\frac{2b}{3}x^{3}<0.$
\end{proof}

\textbf{Remark.} The mere existence of closed orbits around the
equilibrium point $\left(  x_{\ast},0\right)  $ could be inferred
from the following theorem: If an equilibrium point $\mathbf{p}$
is a center for the linearised system and all trajectories are
symmetric with respect to the $x$ axis, then $\mathbf{p}$ is also
a center for the nonlinear system (\ref{2dim}) (cf. \cite{perko}
Theorem 6, page 141). In the above proof we also determine the
subset of the phase space which contains all periodic orbits.

Using all this information we may sketch the phase portrait of the
system (Fig. 3). \begin{figure}[h]
\begin{center}
\includegraphics[scale=0.9]{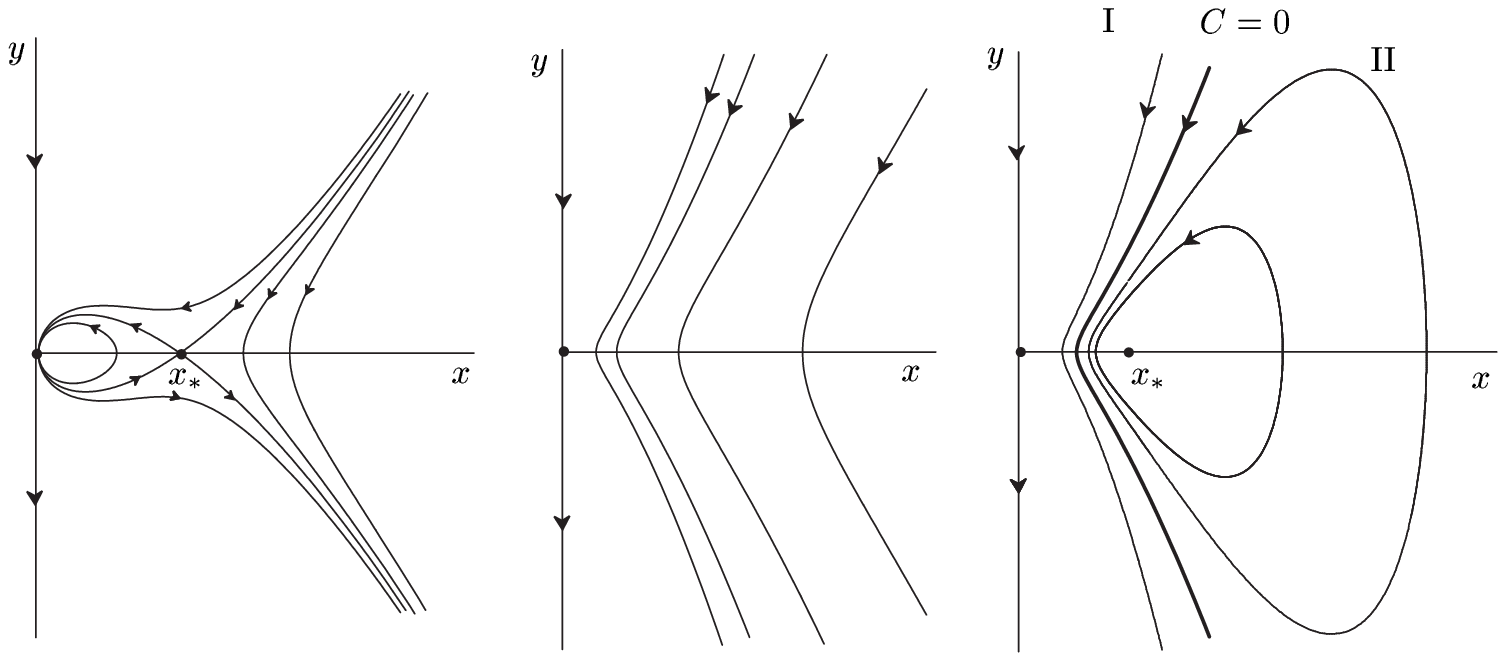}
\end{center}
\caption{Phase portrait of (\ref{2dim}) for $\gamma<\frac{2}{3}$, $\frac{2}%
{3}<\gamma<1$ and $\gamma>1.$}%
\label{fig3}%
\end{figure}

For $\gamma<2/3$ the homoclinic curves to the origin as well as
the curves approaching the origin indicate that an initially
expanding closed universe may avoid recollapse. This result is
also valid in GR when matter fields violate the strong energy
condition \cite{bgt}. For $\gamma\in(\frac{2}{3},1]$ every
solution curve becomes unbounded and we may interpret this as an
indication that this universe recollapses.

The range $(1,2]$ for $\gamma$ is the more interesting because of
the periodic orbits in region II. The phase portrait in Fig. 3 may
lead to the conclusion that the periodic orbits are far from the
origin. However, the position of the cycles in the phase space
depends on the constants $b$ in (\ref{2dim}) and $C$ in (\ref{xy})
and therefore, for suitable values of $b$ and $C,$ there exist
periodic orbits arbitrary close to the origin. Note that the
periodic solutions of (\ref{2dim}) induce periodicity to the full
four-dimensional system (\ref{cyli}) or (\ref{normalf}). In fact,
if $x\left(  t\right)  $ and $y\left(  t\right)  $ are periodic
solutions, then (\ref{r-equ}) implies that the solutions
\[
y_{1}\left(  t\right)  =r\left(  t\right)  \cos\left(
t+\theta_{0}\right) ,\ \ \ y_{2}\left(  t\right)  =r\left(
t\right)  \sin\left(  t+\theta _{0}\right)
\]
oscillate in the $y_{1}-y_{2}$ plane with a periodic amplitude
$r(t).$ Note also that the periodic motion in the $x-y$ plane is
independent from the rotation of $r$ in the $y_{1}-y_{2}$ plane.

Obviously we cannot assign a physical meaning to the new variables
$\left( y_{1},y_{2},x,y\right)  ,$ since the transformations
(\ref{line}) and (\ref{nonl1}) have \textquotedblleft
mixed\textquotedblright\ the original variables of (\ref{full}) in
a nontrivial way. However, the periodic character of the solutions
of (\ref{normalf}) whatever the physical meaning of the variables
be, has the following interpretation. \emph{Close to the
equilibrium of the original system (\ref{full}), there exist
periodic solutions}. This result is in agreement with \cite{cds}
where it is shown that for Lagrangians $R+\beta R^{m},$ bounces of
closed models are allowed for every integer value of $m$.

\textbf{Remark}. As mentioned in Section 2 numerical experiments
show that the solutions of the system have oscillatory behavior.
This property is intuitively evident by looking at the harmonic
oscillator, equation (\ref{trac}). Using the rescalings of the
dynamical variables along with the scaling
$\rho\rightarrow\beta\rho,$ equation (\ref{trac}) becomes
\[
\ddot{R}+3H\dot{R}+R=\frac{\left(  4-3\gamma\right)  }{2}\rho,
\]
which is the equation of a forced, damped harmonic oscillator with
unit angular frequency. Qualitative arguments supported by
numerical solutions indicate that oscillatory motion, possibly
slightly damped, is essentially independent of $k$ and $\gamma$
and\ becomes the late time behaviour. This is revealed in the
normal forms of the systems, (\ref{cylifl}) and (\ref{cyli}),
where the $\theta$ equation shows oscillatory motion with unit
angular frequency. We emphasize again that this motion is
independent from the periodic motion in the $x-y$ plane. The
normal form analysis reveals both kinds of oscillatory behaviour
and distinguishes the models which actually exhibit undamped
oscillations at late time.

\section{Discussion}

We analysed the qualitative behaviour of flat and positively
curved FRW models filled with ordinary matter described by a
perfect fluid in the Jordan frame of the $R+\beta R^{2}$ theory.
We have shown that initially expanding flat models close to the
equilibrium $H=0,R=0,\rho=0$ are ever expanding and asymptotically
approach flat empty spacetime. Therefore, the late time behaviour
of flat models is a common property of quadratic gravity and GR.
Closed models can avoid recollapse for $\gamma<2/3$, but not for
$\gamma$ in the range $\left[  \frac{2}{3},1\right]  ,$ thereby
also behaving similarly to the corresponding general-relativistic
cosmologies.

The interesting feature is the existence of periodic solutions
near the
origin for $\gamma >1$. This is not revealed in the Einstein frame (see \cite%
{miri2}), possibly because in that investigation, the scalar field
related to the conformal transformation is not coupled to matter,
i.e., the matter Lagrangian is added after performing the
conformal transformation. Were the two frames physically
equivalent then, a naive physical explanation of the cyclic
behavior could rely on the scalar field which behaves like a
\textquotedblleft cosmological constant\textquotedblright\ in the
high curvature limit. The perfect fluid which is also present
dominates in the low curvature regime allowing for a recollapse,
but then the effective cosmological constant induces a bounce in
the high-curvature regime.

The periodic solutions imply that an initially expanding closed
universe can avoid recollapse through an infinite sequence of
successive expansions and contractions. The oscillatory open model
proposed by Steinhardt and Turok \cite{sttu} has renewed interest
in cyclic universes. However, observations do not exclude $\Omega
_{\mathrm{total}}$ to be slightly larger than one \cite{jaffe}.
Oscillatory closed models were considered in the context of Loop
Quantum Cosmology \cite{lids-etal} and oscillatory (but not
periodic) solutions in GR were found in \cite{clba} for closed
models containing radiation and dust or scalar field. In Fig. 3
the basin of all periodic trajectories of the (\ref{2dim}) is the
domain on the right of the $C=0$ curve and since it is an open
subset of the phase space, we conclude that there is enough room
in the set of initial data of (\ref{full}) which lead to an
oscillating scale factor. The $R+\beta R^{2}$ theory has offered a
successful inflationary model, but whether it is capable to
explain the acceleration of the universe is an open question that
needs to be studied in more detail. In particular, the observed
slow acceleration must be related to the periods of the closed
curves of (\ref{2dim}).

The normal form theory is a powerful method for determining the
qualitative behaviour of a dynamical system near a nonhyperbolic
equilibrium point, but does not give any information about the
structure of the solutions far from this equilibrium. Our results
are based on an analysis of the behaviour of the dynamical system
(\ref{full}) only near the equilibrium solutions. The geometry of
the trajectories of a four-dimensional dynamical system may be
quite complicated, e.g. strange attractors may be present. For the
system (\ref{full}) the whole picture may come in view only from
the investigation of the global structure of the solutions. The
study of this question is an interesting challenge for
mathematical relativity.

\section*{Acknowledgements}

I thank Spiros Cotsakis and Alan Rendall for useful comments. I am
grateful to the referee for comments from the physical
perspective.

\renewcommand{\theequation}{A.\arabic
{equation}}

\section*{Appendix: Normal form of (\ref{full})}

Following the usual algorithm, we use the matrix which transforms
the linear
part of (\ref{full}) into Jordan canonical form and define new variables by%
\begin{equation}
y_{1}=x_{1},\ \ y_{2}=-x_{2},\ \ x=x,\ \ y=H+2x_{2},\label{line}%
\end{equation}
so that (\ref{full}) becomes%
\begin{eqnarray*}
&&\left[
\begin{array}
[c]{c}%
\dot{y}_{1}\\ \dot{y}_{2}\\ \dot{x}\\
\dot{y}%
\end{array}
\right]=\left[
\begin{array}
[c]{cccc}%
0 & -1 & 0 & 0\\ 1 & 0 & 0 & 0\\ 0 & 0 & 0 & 0\\ 0 & 0 & 0 & 0
\end{array}
\right]  \left[
\begin{array}
[c]{c}%
y_{1}\\ y_{2}\\ x\\ y
\end{array}
\right]+ \\&& \left[
\begin{array}
[c]{c}%
0\\
\left(  4-3\gamma\right)  y_{1}^{2}-\left(  3\gamma+2\right)  y_{2}^{2}%
-\frac{4-3\gamma}{4}\left(  kx^{2}+y^{2}\right)  -3yy_{2}-\left(
4-3\gamma\right)  Z_{3}\\ -2xy_{2}-xy\\
-2\left(  4-3\gamma\right)  y_{1}^{2}+2\left(  3\gamma-2\right)  y_{2}%
^{2}-\frac{3\gamma-2}{2}kx^{2}-\frac{3\gamma}{2}y^{2}-2yy_{2}+2\left(
4-3\gamma\right)  Z_{3}%
\end{array}
\right]  ,
\end{eqnarray*}
with $Z_{3}=y_{1}\left[  kx^{2}+\left(  2y_{2}+y\right)
^{2}\right]  .$ Under
the non-linear change of variables%
\begin{eqnarray}
y_{1}&\rightarrow& y_{1}+3\gamma y_{1}^{2}+\left( 3\gamma-2\right)
y_{2}^{2}+\frac{4-3\gamma}{4}\left(  x^{2}+y^{2}\right)  +\frac{3}{4}%
y_{2}y,\nonumber\\ y_{2}&\rightarrow&
y_{2}+4y_{1}y_{2}+\frac{3}{4}y_{1}y,\label{nonl1}\\ x&\rightarrow&
x+2y_{1}x,\nonumber\\ y&\rightarrow&
y-2y_{1}y_{2}+2y_{1}y,\nonumber
\end{eqnarray}
and keeping only terms up to second order, the system transforms to%

\begin{eqnarray}
\dot{y}_{1}&=&-y_{2}-\frac{3}{2}y_{1}y+\mathcal{O}\left( 3\right)
,\nonumber\\
\dot{y}_{2}&=&y_{1}-\frac{3}{2}y_{2}y+\mathcal{O}\left(  3\right)
,\label{normalf}\\ \dot{x}&=&-xy+\mathcal{O}\left(  3\right)
,\nonumber\\ \dot{y}&=&6\left(  \gamma-1\right)  \left(
y_{1}^{2}+y_{2}^{2}\right)
-\frac{3\gamma-2}{2}x^{2}-\frac{3\gamma}{2}y^{2}+\mathcal{O}\left(
3\right) .\nonumber
\end{eqnarray}


\begin{thebibliography}{99}                                                                                               %


\bibitem {lno}Lidsey J.E., Nojiri, S., Odintsov, S.D.: JHEP \textbf{0206} 026
(2002); Lidsey, J.E., Nunes, N.J.: Phys. Rev. D \textbf{67} 103510 (2003)

\bibitem {nos}Nojiri, S., Odintsov, S.D., Sasaki, M.: Phys. Rev. D \textbf{71}
123509 (2005)

\bibitem {gave}Gasperini, M., Veneziano, G.: Phys. Rept. \textbf{373} 1 (2003)

\bibitem {stel}Stelle, K.:\textit{ }Phys. Rev.\textbf{\ }D\textbf{ 16} 953 (1977)

\bibitem {star}Starobinsky, A.: Phys. Lett\emph{.} B \textbf{91} 99 (1980)

\bibitem {cdtt}Carroll, S.M., Duvvuri, V., Trodden, M., Turner, M.S.:\emph{
}Phys. Rev. D \textbf{70} 043528 (2004)

\bibitem {cct2}Capozziello, S., Carloni, S., Troisi, A.: Preprint
astro-ph/0303041 (2003)

\bibitem {caretal}Carroll, S.M., De Felice, A., Duvvuri, V., Easson, D.A.,
Trodden, M., Turner, M.S.: Phys. Rev. D\textbf{ 71} 063513 (2005)

\bibitem {nood}Nojiri, S., Odintsov, S.D.: Int. J. Geom. Meth. Mod. Phys.
\textbf{4} 115 (2007)

\bibitem {cno}Capozziello, S., Nojiri, S., Odintsov, S.D.: Phys. Lett. B
\textbf{634} 93 (2006); Capozziello, S., Nojiri, S., Odintsov, S.D. and
Troisi, A.: Phys. Lett. B \textbf{639} 135 (2006)

\bibitem {amen1}Amendola, L., Polarski, D., Tsujikawa, S.: Phys. Rev. Lett.
\textbf{98} 131302 (2007); Amendola, L., Polarski, D., Tsujikawa, S.: Int. J.
Mod. Phys. D\textbf{16} 1555 (2007)

\bibitem {amen2}Amendola, L., Gannouji, R., Polarski, D., Tsujikawa, S.: Phys.
Rev. D \textbf{75} 083504 (2007)

\bibitem {cafr}Capozziello, S., Francaviglia, M.: Gen. Relativ. Gravit.
\textbf{40}, 357 (2008)

\bibitem {cct1}Capozziello, S., Cardone, V.F., Troisi, A.: Phys. Rev. D
\textbf{71} 043503 (2005)

\bibitem {soko}Soko\l owski, L.M.: Class. Quantum Grav. \textbf{24} 3391 (2007)

\bibitem {miri1}Miritzis, J.: Preprint\emph{ }gr-qc/0609025 (2006)

\bibitem {bahe}Barrow, J.D., Hervik, S.: Phys. Rev. D \textbf{74} 124017 (2006)

\bibitem {nood1}Nojiri, S., Odintsov, S.D.: Phys. Rev. D \textbf{74} 086005 (2006)

\bibitem {bran}Brans, C.H.: Class. Quantum Grav. \textbf{5} L197 (1998);
Ferraris, M., Francaviglia, M., Magnano, G.: Class. Quantum Grav. \textbf{7}
261 (1990); Cotsakis, S.: Phys. Rev.\textbf{\ }D \textbf{47} 1437 (1993);
Erratum Phys. Rev.\textbf{\ }D \textbf{49} 1145 (1994)

\bibitem {maso}Magnano, G., Soko\l owski, L.M.: Phys. Rev. D\textbf{ 50} 5039
(1994); Faraoni, V., Gunzig, E., Nardone, P.: Fund. Cosmic Phys. \textbf{20,}
121 (1999); Faraoni, V., Nadeau, S.: Phys. Rev. D \textbf{75} 023501 (2007)

\bibitem {baot}Barrow, J.D., Ottewill, A.: J. Phys. A \textbf{16 }35 (1983)

\bibitem {olmo}Olmo, G.: Phys. Rev. D \textbf{75} 023511 (2007)

\bibitem {wael}Wainwright, J., Ellis, G.F.R.: Dynamical Systems in Cosmology.
Cambridge University Press (1997)

\bibitem {bami}Barrow, J.D., Middleton, J.:\emph{ }Preprint\emph{
}gr-qc/0702098 (2007)

\bibitem {cofl}Cotsakis, S., Flessas, G.: Phys. Rev. D \textbf{48} 3577
(1993); Cotsakis, S., Flessas, G.: Phys. Rev. D \textbf{51} 4160 (1995)

\bibitem {berk}Berkin, A.: Phys. Rev. D\textbf{ 42} 1016 (1990)

\bibitem {ctd}Carloni, S., Troisi, A., Dunsby, P.K.S.: Preprint\emph{
}arXiv/0706.0452 (2007)

\bibitem {gld}Goheer, N., Leach, J.A., Dunsby, P.K.S.: Class. Quantum Grav.
\textbf{25} 035013 (2008)

\bibitem {perko}Perko, L.: Differential Equations and Dynamical Systems.
Springer-Verlag (2001)

\bibitem {guho}Guckenheimer, J., Holmes, P.: Nonlinear Oscillations, Dynamical
Systems and Bifurcations of Vector Fields. Springer-Verlag, New York (1983)

\bibitem {takens}Takens, F.: Publ. Math. IHES \textbf{43} 47 (1974)

\bibitem {miri2}Miritzis, J.: J. Math. Phys. \textbf{46} 082502 (2005)

\bibitem {bgt}Barrow, J.D., Galloway, G., Tipler, F.: Mon. Not. R. Astr. Soc.
\textbf{223 }835 (1986)

\bibitem {cds}Carloni, S., Dunsby, P.K.S., Solomons, D.: Class. Quantum Grav.
\textbf{23} 1913 (2006)

\bibitem {sttu}Steinhardt, P.J., Turok, N.: Science \textbf{296} 1436 (2002)

\bibitem {jaffe}Jaffe, A.H. et al.: Phys. Rev. Lett. \textbf{86} 3475 (2001)

\bibitem {lids-etal}Lidsey, J.E., Mulryne, D.J., Nunes, N.J., Tavakol, R.:
Phys. Rev. D \textbf{70} 063521 (2004)

\bibitem {clba}Clifton, T., Barrow, J.D.: Phys. Rev. D \textbf{75} 043515 (2007)
\end{thebibliography}
\end{document}